\documentclass[preprint,5p,twocolumn]{elsarticle}
\usepackage{hyperref}
\usepackage{lineno}
\usepackage{upgreek}
\usepackage{float}
\usepackage{bm}
\usepackage{xcolor}
\usepackage{ulem}
\usepackage{amsmath} 
\usepackage{graphicx}
\usepackage[english]{babel}  
\journal{arXiv}
\begin{document}
 
\title{A comparative study of solvent dynamics in Choline Bromide aqueous solution using combination of neutron scattering technique and molecular dynamics simulation}     
\author[fn1]{Debsindhu Bhowmik\corref{mycorrespondingauthor1}}
\cortext[mycorrespondingauthor1]{Corresponding author}
\ead{bhowmikd@ornl.gov}   
\address[fn1]{Computational Science and Engineering Division, Oak Ridge National Laboratory, Oak Ridge,  Tennessee 37831, USA.}   
\date{\today}

\begin{abstract}

A comparative study between Choline and Tetra-methyl ammonium bromide is presented here. Choline which is a crucial component for our dietary requirements causes many diseases if there is deficiency. We used combined approach of all-atom molecular dynamics simulation coupled with neutron scattering technique to study mainly the solvent behavior in this study. There is follow up work where we discussed about the solute dynamical behavior.               

\end{abstract}

\maketitle


\section{\label{Intro} Introduction}      

\begin{figure}[!htbp]
  \begin{center}
    \includegraphics[width=0.45\textwidth,angle=-180]{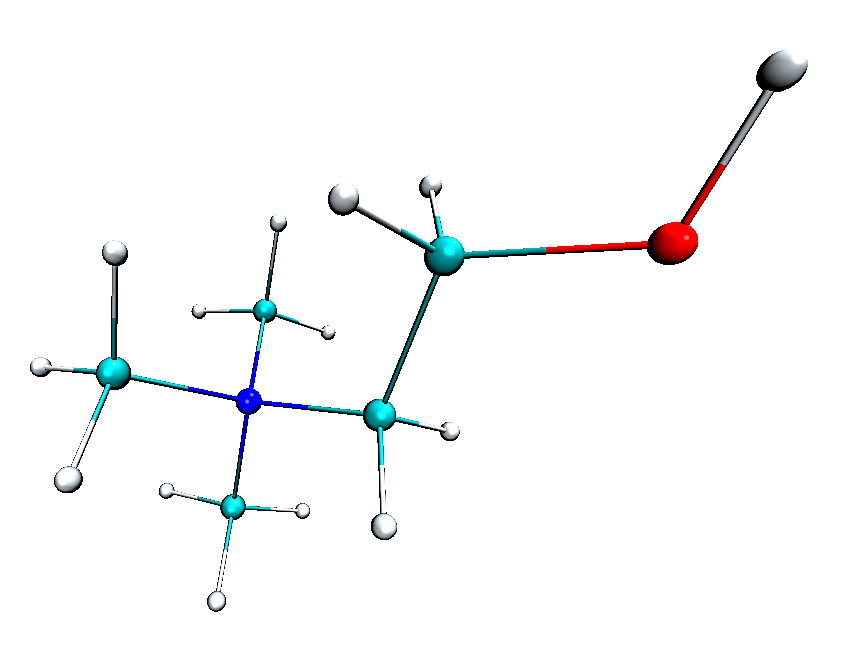} 
  \end{center}
  \bf\caption {Choline$^{+}$ cation. green: carbon, blue: nitrogen, white: hydrogen and red: oxygen atom.}
  \label{fig:choline}
\end{figure}

One very important and must dietary component for us is the Choline (figure~\ref{fig:choline}). This is an indispensable nutrient. Choline is water soluble and could be found in human milk in free Choline form or in phosphocholine and glycerophosphocholine as primary structures. This belongs to the family of Tetra-alkyl ammonium (TAA) family containing \texttt{quaternary amine}. This is important for many of the bio-physical processes such as signaling functions of cell membranes, lipid transport or metabolism etc. Choline is also important during pregnancy and development of fetus. Lack of Choline in body could lead to diseases related to liver or neurological disorders.              

We will study about the solvent behavior of Choline in comparison to TMA (Tetra-methyl ammonium), the smallest in the TAA family with Bromide as anion in the aqueous solution. The structural difference between Choline and TMA is one additional -CH$_2$(OH) group attached to one arm of TMA. Further the solvent dynamics including the separation of translational and rotational part of the hydration water molecules around the molecule will be discussed. We will primarily make use of the all-atom Molecular Dynamics (MD) simulation for elucidating the structure and the dynamics. The experiments will be carried out by scattering techniques mainly with Time of Flights spectrometers.      
\section{\label{methods} Materials and Methods}\ 
\subsection{\label{simudetails} Simulation Details}\ 
\subsubsection{choosing of models}    
The details of the MD simulation is discussed in this work~\cite{Bhowmik6} where the classical molecular dynamics (MD) simulations (by DL POLY 2.18~\cite{Smith07}) is carried out with aqueous CholineBr, TMABr, NaBr, KBr and CsBr solutions. We have considered all atom (i.e. explicit N, C, H atoms and also O for Choline), flexible (i.e. bond stretch, bond bending, dihedral interaction) and non-polarizable model. In these cases we calculated the atomic charges by Hartree-Fock method considering nonpolarizable force fields with the proper modification involving the AMBER supported Antechamber routine~\cite{Heyda10}. The rest of the parameters for interactions are imported from Generalized Amber Force Field (GAFF)~\cite{AMBER_10}. A brief summary of the cationic atom charges and force field parameters for TMA$^+$ are presented in table~\ref{tab:TMA_charge},~\ref{tab:choline_charge},~\ref{tab:TMA_FF} and~\ref{tab:choline_FF}. The atomic charges or force-field parameters for the other systems like sodium, potassium or bromide are considered from earlier literatures~\cite{Koneshan98_102}~\cite{Horinek09}~\cite{Lee96} ~\cite{Joung08}~\cite{Markovich96_7}. Selecting proper water model remains always one of the most difficult tasks. We have selected SPC/E water model due to its ability to faithful reproduction of experimental structural and dynamical characteristics in vast range of pressure or temperature~\cite{Brodholt93}. The SPC/E model which is a rigid model where O-H bond = 1.0~\AA~\, H-O-H angle = 109$^\circ$ and charges = +0.424e (hydrogens) and -0.848e (oxygens) respectively~\cite{Berendsen87_91}. Note that this is an extended version of the SPC model~\cite{Guillot02} where an extra correction on self polarazibility is included. The Lorentz-Berthelot mixing rule is used for L-J parameters to define non-bonded force fields along with Coulombic potentials.              

\subsubsection{performing simulation}    
\begin{figure}[!htbp]
  \begin{center}
    \includegraphics[width=0.45\textwidth,angle=0]{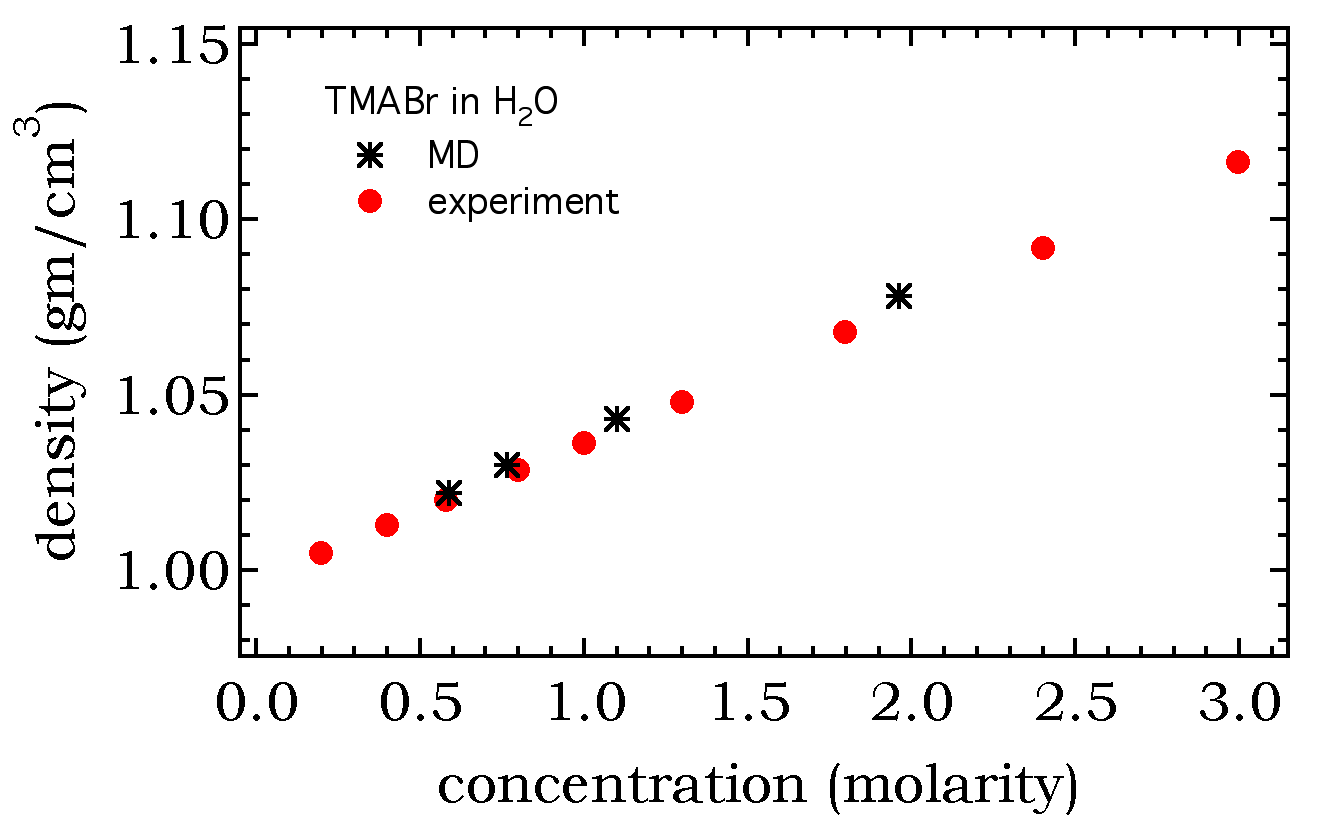} 
  \end{center}
   \bf\caption {Comparison of density for aqueous TMABr solution: extracted from the MD simulation with experiment~\cite{Buchner02}.}
  \label{fig:TMABr_density}
\end{figure}

\begin{figure}[!htbp]
  \begin{center}
    \includegraphics[width=0.45\textwidth,angle=0]{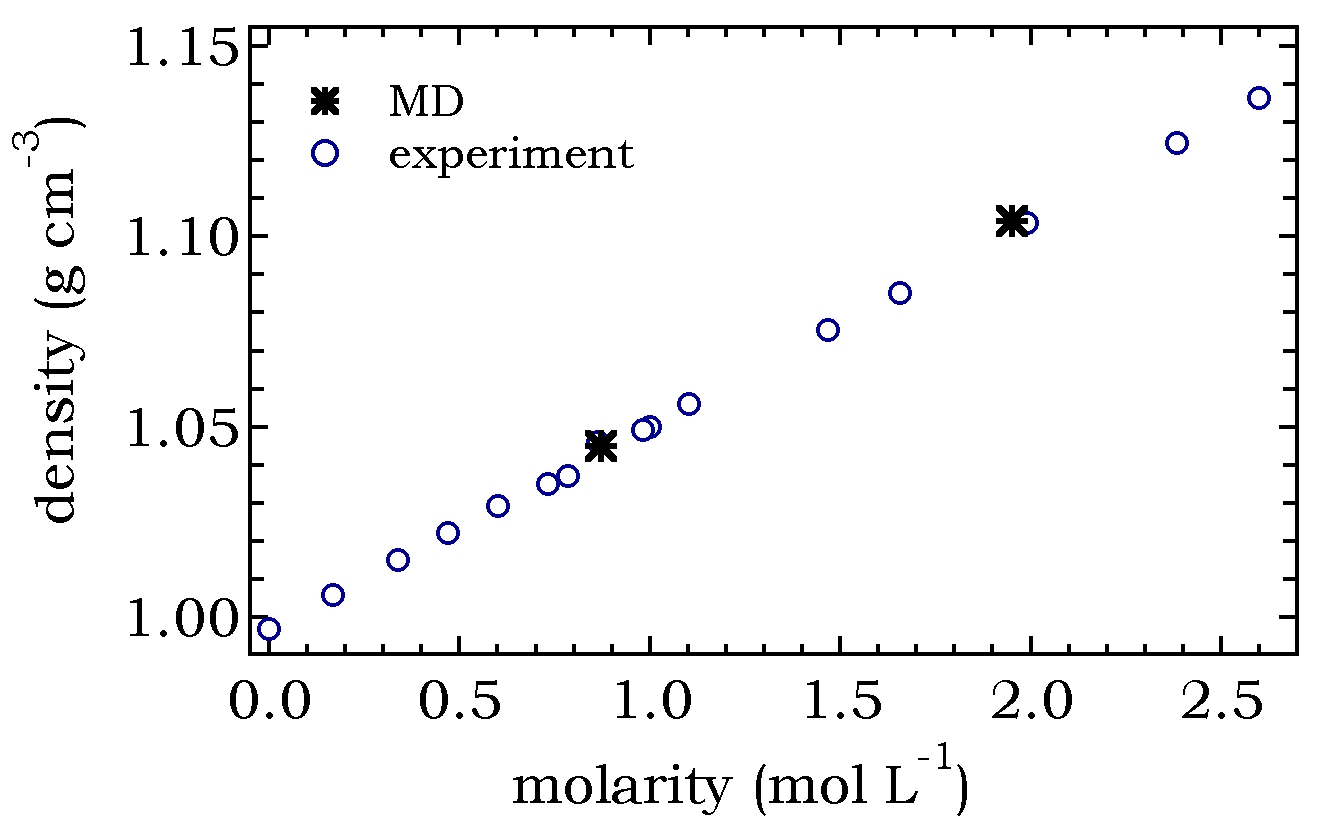} 
  \end{center}
   \bf\caption {Comparison of density for aqueous CholineBr solution: extracted from the MD simulation with experiment.}
  \label{fig:choline_density}
\end{figure}

Prior to the commencement of the simulation one simulation box was built that is constituted with the proper geometry. This is carried out in accordance with the bond length, valence or dihedral angles computed experimentally. In the next step one simulation box is built that satisfies the collective volume of the desired number of cations, anions and solvent molecules. Once the solvent molecules filled the rest of the volumes with random placements and orientation we placed cations and anions in the box. We took care of the fact that the state of the simulation box is not far from the equilibrium state. This completes the building of the simulation box required to start the simulation. Next we commence the simulation in NPT ensemble till the energies - both potential and the kinetic - are stable with temperature, volume and pressure stay steady. The system density now satisfies the experimental results. For the simulation we have used three-dimensional periodic boundary condition with cut-off radius specified for short range interaction equal to half-the-box-size. Three dimensional Ewald sum takes care of the long range interactions and SHAKE algorithm is used for rigid model of SPC/E water. First the equilibration with initial configurations are carried out in NPT ensembles and after that the production run is carried out in NVE ensemble with timestep of 1fs for 3.4ns. The saved frames are every 0.1ps generating total of 34$\times$10$^{3}$ frames. The nMoldyn~\cite{nMOLDYN} is then used to analyze the data. We verified that our simulations carried out in different systems with different concentrations do not vary more than 0.2\% compared to the experimental values. Comparisons between CholineBr and TMABr aqueous solutions are presented in figure~\ref{fig:TMABr_density} and~\ref{fig:choline_density}.           

\begin{table}[!htbp]    
\scriptsize      
\begin{tabular}{|l|ccc|}
	\hline
part of molecule & atom charge (e) & from bold part & \\
 & N & C & H \\
	\hline	
-\bf{N}- & 0.096521 & & \\
-N-$\bf{CH_{3}}$ & & -0.165381 & 0.130417 \\
	\hline
\end{tabular}
 \bf\caption {atomic charge distribution: TMA$^+$.}
 \label{tab:TMA_charge}
\end{table}

\begin{table}[!htbp]    
\scriptsize    
\begin{tabular}{|l|p{0.8cm}p{0.8cm}p{0.8cm}p{0.8cm}|}
	\hline
part of molecule & charge & (e) & & \\
 & N & C & H & O \\
	\hline	
-\bf{N}- & 0.0222 & & & \\
-N-$\bf{CH_{3}}$ & & -0.142 & 0.124 & \\
-N-$\bf{CH_{2}}$- & & -0.050 & 0.134 & \\
-N-$CH_{2}-\bf{CH_{2}}$- & & -0.183 & 0.043 & \\
-N-$CH_{2}-CH_{2}-\bf{OH}$- & & & 0.473 & -0.669\\
	\hline
\end{tabular}
 \bf\caption {atomic charge distribution: Choline$^+$.}
 \label{tab:choline_charge}
\end{table} 

\begin{table}[!htbp]    
\begin{tabular}{|c|cc|}
	\hline
bond elongation & energy & length \\
harmonic parameters & (kcal/mol/\AA$^{2}$) & (\AA) \\
	\hline	
C-H$_{N}$ & 240 & 1.090 \\
C-N & 367 & 1.471 \\
	\hline
\end{tabular}
\begin{tabular}{|c|cc|}
	\hline
bond bending & energy & angle \\
harmonic parameters & (kcal/mol/rad$^{2}$) & (degree) \\
	\hline
H$_{N}$-C-H$_{N}$ & 35 & 109.5\\
H$_{N}$-C-N & 50 & 109.5 \\
C-N-C & 50 & 109.5 \\
	\hline
\end{tabular}
\begin{tabular}{|c|cc|}
	\hline
dihedral interaction & energy & angle \\
 & (kcal/mol) & (degree) \\
	\hline
X-C-N-X & 0.15 & 0.0 \\
	\hline
\end{tabular}
\begin{tabular}{|c|cc|}
	\hline
L-J & $\epsilon$ & $\sigma$ \\
parameters & kcal/mol & \AA \\
	\hline
H$_{N}$ & 0.0157 & 1.100\\
C & 0.1094 & 1.900 \\
N & 0.1700 & 1.8240 \\
	\hline
\end{tabular}
\bf\caption {TMA$^+$ Force Field.} {Force field parameters for TMA$^+$ atoms (H$_{N}$ for the hydrogens of the carbon attached to the central N).}
\label{tab:TMA_FF}
\end{table}

\begin{table}[!htbp]    
\begin{tabular}{|c|cc|}
	\hline
bond elongation & energy & length \\
harmonic parameters & (kcal/mol/\AA$^{2}$) & (\AA) \\
	\hline	
C-H$_{N}$ & 240 & 1.090 \\
C-H$_{C}$ & 340 & 1.090 \\
C-C & 310 & 1.526 \\
C-N & 367 & 1.471 \\
C-O$_{OH}$ & 320 & 1.410 \\
C-H$_{OH}$ & 553 & 0.960 \\
	\hline
\end{tabular}
\begin{tabular}{|c|cc|}
	\hline
bond bending & energy & angle \\
harmonic parameters & (kcal/mol/rad$^{2}$) & (degree) \\
	\hline
H$_{N}$-C-H$_{N}$ & 35 & 109.5\\
H$_{C}$-C-H$_{C}$ & 35 & 109.5\\
C-C-H$_{C}$ & 50 & 109.5\\
C-C-H$_{N}$ & 50 & 109.5\\
H$_{C}$-C-O$_{OH}$ & 50 & 109.5\\
C-C-N & 80 & 111.2\\
H$_{N}$-C-N & 50 & 109.5\\
C-N-C & 50 & 109.5\\
C-C-O$_{OH}$ & 50 & 109.5\\
C-O$_{OH}$-H$_{OH}$ & 35 & 109.5\\
	\hline
\end{tabular}
\begin{tabular}{|c|cc|}
	\hline
dihedral interaction & energy & angle \\
 & (kcal/mol) & (degree) \\
	\hline
X-C-N-X & 0.15 & 0.0 \\
X-C-C-X & 0.15 & 0.0 \\
X-C-O$_{OH}$-X & 0.15 & 0.0 \\
	\hline
\end{tabular}
\begin{tabular}{|c|cc|}
	\hline
L-J & $\epsilon$ & $\sigma$ \\
parameters & kcal/mol & \AA \\
	\hline
H$_{N}$ & 0.0157 & 1.387\\
H$_{N}$ & 0.0157 & 1.100\\
H$_{OH}$ & 0.0000 & 0.0000\\
O$_{OH}$ & 0.2104 & 1.7210\\
C & 0.1094 & 1.900 \\
N & 0.1700 & 1.8240 \\
	\hline
\end{tabular}
\bf\caption {Choline$^+$ Force Field.} {Force field parameters for Choline$^+$ atoms (H$_{N}$ for the hydrogens of the carbon attached to the central N).}
\label{tab:choline_FF}
\end{table}   

Here all of the MD simulations are carried out with H$_2$O as the solvent components. For the neutron related analysis the deuterium scattering lengths are taken for the H atoms in water for the analysis. This is performed to reproduce the experiments with D$_2$O solvent.     

\subsection{\label{expdetails} Experimental Details}\   
The experimental set-up for the neutron experiment is as follows. 

\subsubsection{\label{sample_preparation} Sample Preparation}    
The sample preparation is similar to the work by Bhowmik et al.~\cite{Bhowmik6}. The protonated samples are bought from Fluka with $>$99\% purity. They are preserved in dry place and in distance or direct contact from sunlight. Before the sample preparation the salts are dried in vacuum for multiple hours to remove any water molecules. Next liquid D$_2$O (Euriso-top, 99.9\%D) or distilled H$_2$O is used to make the proper concentration ration between salt and solvent. Nitrogen gas is then used to store the remaining salts. The deuteration stage 's crucial 'cause of the uniqueness of the neutron scattering techniques. The inexchangibility of the hydrogen atoms in the cations with the solvent keeps the solute and solvent character intact.                   

\subsubsection{\label{experimental_setup} Experimental Setup}   
The Time of Flight (ToF) technique is used for the neutron scattering experiments. The details or related information could be found in ref~\cite{Bhowmik, Bhowmik1, Bhowmik2, Bhowmik3, Bhowmik4, Bhowmik5, Bhowmik6, Bhowmik7, Bhowmik8, Bhowmik9, Bhowmik10, Bhowmik11, Bhowmik12, Bhowmik13, Bhowmik14, Bhowmik15}. For the experiments 0.2 mm thick flat quartz cell is used in the MIBEMOL spectrometer in LLB-Orphee reactor with resolution of 50 $\mu$eV (HWHM) under 6 \AA incident beam. The vanadium 's used fr experimental resolution. The $Q$ range covers from 0.49 \AA$^{-1}$ to 1.97 \AA$^{-1}$. We made sure that no sample is lost during the experiment by weighing before and after the experiments.                      

\section{\label{results} Results}         
\subsection{solvent structure}    

Here we discuss the solvent structure and dynamics. Like before we concentrate on aqueous CholineBr solution along with TMABr and NaBr with $x_m$=1:56. And at the end, the results for higher concentration (with $x_m$=1:22) for NaBr, KBr, CsBr and TMABr solutions will be discussed. We will start by simulation results and then the experimental results will be presented. 

\begin{figure}[!htbp]
\begin{center}
\includegraphics[width=0.45\textwidth,angle=0] {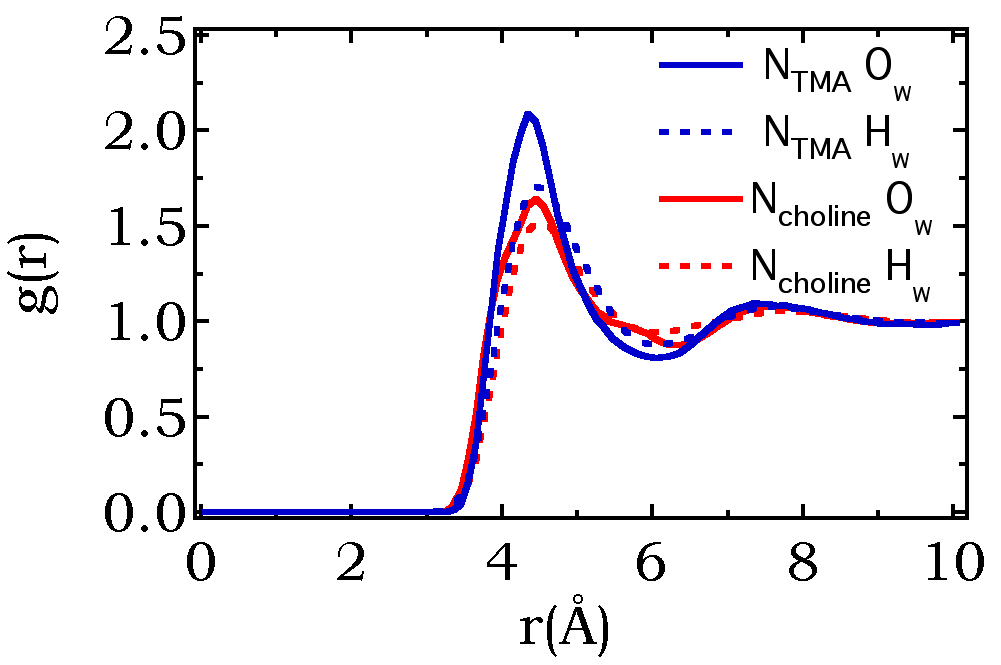}
\vspace{0.1cm}
\bf\caption{Cation-water Radial Distribution Function for aqueous TMABr and CholineBr solutions with $x_m$=1:56.}
\label{TMA_choline_N-H2O}
\end{center}
\end{figure}

The Choline hydration structure is similar to TMA$^+$ except where we observe that the first peak of cation-water (oxygen) of choline$^+$ is less intense than TMA$^+$ and the g(N$_{choline}$O$_{W}$) and g(N$_{choline}$H$_{W}$) are closer to each other than for TMA$^+$ (figure \ref{TMA_choline_N-H2O}). 

\subsection{solvent dynamics}   
\subsubsection{solvent dynamics: Translational}    

The average solvent D$_{tr}$ is also estimated by MD simulation. If we focus on aqueous TMABr and NaBr solution with $x_m$=1:56, we see that the average water diffusion coefficient is 1.5 higher than for aqueous TBABr at the same concentration. At a concentration of $x_m$=1:22 (TMABr and NaBr) average solvent $D_{tr}$ is further reduced. Due to unavailability of deuterated CholineBr and TBABr, experimental solvent diffusion can not not be estimated. The simulation suggests that CholineBr reduces average solvent dynamics more than TMABr or simple salts like NaBr when concentration is increased from $x_m$=1:56 to $x_m$=1:22. The result is summarized in table \ref{tab:salt_diff}.

\begin{table} 
\scriptsize  
\begin{tabular}{|c|l|cc|}
	\hline
 & & $D_{tr}$ of solvent water & in (10$^{-9}$ m$^2$s$^{-1}$) \\ 
 	\hline
concentration & Aqueous & exp & sim \\
 & solution of & & \\
        \hline	
$x_m$=1:112 & TBABr & & (1.78$\pm$0.01) \\
        \hline	        
$x_m$=1:56 & TBABr & (1.40$\pm$0.4) & (1.26$\pm$0.01) \\
  & TMABr & (1.83$\pm$0.10) & (1.98$\pm$0.01) \\
& CholineBr & & (2.10$\pm$0.01) \\
& NaBr & & (2.00$\pm$0.01) \\
      \hline
$x_m$=1:22 & TMABr & (1.31$\pm$0.03) & (1.63$\pm$0.01) \\
& CholineBr & & (1.44$\pm$0.01) \\
& NaBr & (1.39$\pm$0.05) & (1.82$\pm$0.01) \\
& KBr & (1.45$\pm$0.05) & \\
& CsBr & (1.43$\pm$0.05) & \\
     \hline	
& bulk H$_2$O & (2.2$\pm$0.1) & (2.35$\pm$0.01) \\     
     \hline
\end{tabular}
 \bf\caption {Extracted translational diffusion coefficient estimated by MD simulation and QENS experiment for average solvent water molecules of different systems [aqueous solution of simple salts (like NaBr, KBr, CsBr), TMABr, CholineBr and TBABr].}
 \label{tab:salt_diff}
\end{table} 

\subsection{\label{dyn_solvent_rot} solvent dynamics: rotational}    
\begin{figure}[!htbp]
\begin{center}
\includegraphics[width=0.45\textwidth,angle=0] {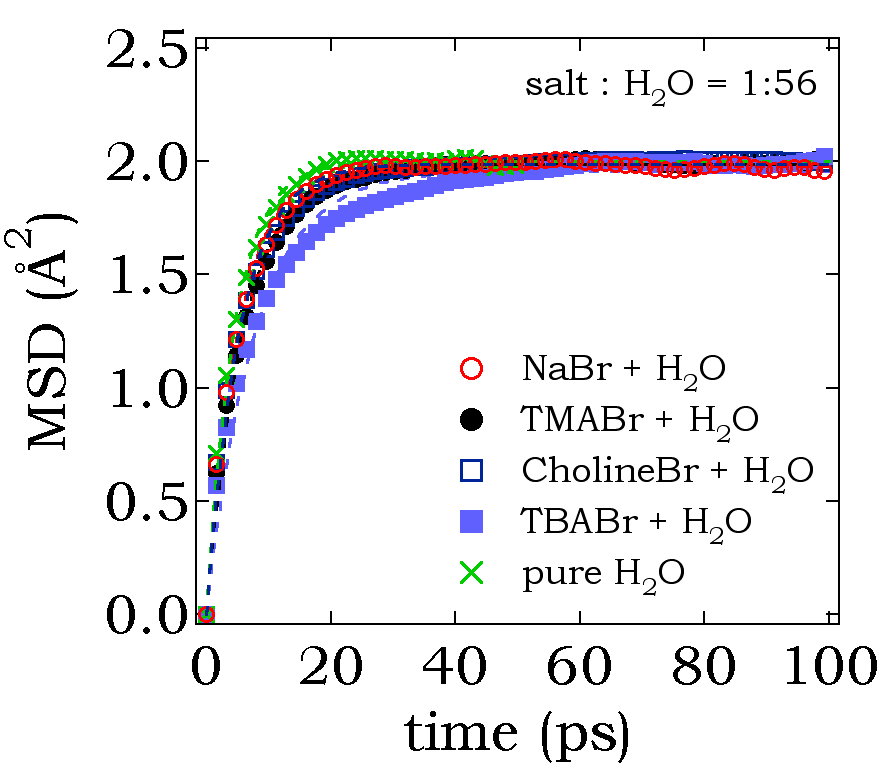}
\vspace{0.1cm}
\bf\caption{MSD of water hydrogen atoms (with modified coordinates relative to its central oxygen atom) are plotted as a function of time to extract estimate the rotation time of water hydrogen atom around central oxygen atom, $\tau_{rot}^{H_{2}O}$. The fitting equation is $2b^2(1-e^{-\frac{t}{\tau_{rot}^{H_{2}O}}})$, where  $b$ is $\sim$1.0\AA$ $ as oxygen-hydrogen bond length.}
\label{rot_H2O}
\end{center} 
\end{figure}

we have modified the individual hydrogen atom (of each H$_2$O molecule) coordinates relative to its central oxygen atoms in the MD simulation trajectory file. This way we can calculate the average water hydrogen atom rotational motion about respective oxygen atom. Next like before we calculate the MSD with the modified water hydrogen atom and fitted with $2b^2(1-e^{-\frac{t}{\tau_{rot}^{H_{2}O}}})$, where $\tau_{rot}^{H_{2}O}$ is the water hydrogen rotational time around oxygen and $b$ is $\sim$1.0\AA$ $ as oxygen-hydrogen bond length (figure \ref{rot_H2O}). We find the $\tau_{rot}^{H_{2}O}$ of solvent water molecules in aqueous solution NaBr, TMABr, CholineBr and TBABr solution are $\sim$1.71ps, $\sim$1.97ps, $\sim$1.79ps and $\sim$2.59ps. However, it is worth noting that rotations of water molecules must be decomposed in their three components. The component that we evaluate here, around the C2 axis, corresponds to a short relaxation time but it is not easily decoupled from the others in dedicated experiments. What makes the situation more complex is the fact that water molecules are bound in average to at least 2 neighbouring molecules and all the motions are coupled at short time scales, in the time domain extending from 1 ps (characteristic lifetime of a hydrogen bond) to 10 ps (the global rotation of the molecule).

\section{ \label{discussion} Discussion}     

In this section we have discussed about the solvent structure and dynamics of aqueous TAABr systems and cholineBr along with normal salts like NaBr, KBr and CsBr at ambient temperature. We show that the hydration structure around TAA salts is different  from normal salts. The water molecules are oriented tangentially around TAA or Choline cation while for NaBr, the oxygen atoms are closer than hydrogen because of the positive nature of the cation. For TBA cation, the water molecules penetrate up to the same distance as TMA. The presence of prominent hydration shell indicates that the cations are placed inside a cage and as the solute size increases the solvation cage becomes weaker. 

The average solvent water dynamics is measured by a combination of TOF and MD simulation. At a concentration of $x_m$=1:56 for aqueous TBABr solution, solvent translational dynamics is decreased by a factor of 2 (for $x_m$=1:112, the factor is 1.3) with respect to bulk. The D$_{tr}$ estimated by TOF for simple salts (NaBr, KBr, CsBr) and TMA at $x_m$=1:22 or 2.5m (1 hydration sphere concentration of TMA cation) shows that solvent dynamics is reduced by a factor of 2 compared to bulk. This indicates that the change in the nature of solute does not significantly alter the solvent dynamics (for TMA or simple salts). But from MD results, it is observed that in case of CholineBr solution, the decrease of solvent dynamics is faster than others. Lastly at any certain concentration ($x_m$=1:56 or 1m for example), the decrease in solvent translational dynamics is larger for the bigger TBA cations (larger size and hydrophobic effect of longer alkyl chain) while smaller TMA cations behaves like normal salts.     

\section{\label{Conclusion} Conclusion}\   

In this work we have discussed the dynamics of solvent in aqueous solutions of TMABr and CholineBr at different concentrations. It is important to understand that the dynamics of water molecules which is measured by the QENS experiment, is an average over several types of water molecules. Some of these water molecules are in the hydration shell or at the vicinity of the ion and others are bulk-like. The difference between concentrations are likely due to the amount of population of each species.            

\section{\label{Acknowledgement} Acknowledgement}\   

D.B. likes to thank all the people who have helped in this side project work during his thesis days. D.B. also likes to point out that many of these results are mentioned in scattered way in his thesis so there was requirement to bring them together.    


\section*{\label{Ref} References}\ 


\end{document}